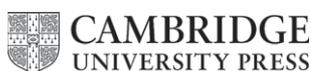

# Why business adoption of quantum and AI technology must be ethical

Christian Hugo Hoffmann[1,2,3] and Frederik F. Flöther[4]

[1]House of Lab Science AG, Hombrechtikon, Switzerland; [2]Technopark Zurich, Zurich, Switzerland; [3]Centre for Ethics of the University of Zurich, Zurich, Switzerland and [4]QuantumBasel, Arlesheim, Switzerland

## Abstract

Artificial intelligence (AI) recently had its "iPhone moment" and adoption has drastically accelerated. Quantum computing appears poised to follow suit over the next years. However, while there has been discourse about how to use AI responsibly, there is still little appreciation and awareness among executives, managers and practitioners about the broader ethical questions and implications raised by the intersection of these emerging technologies. In this article, it is highlighted why quantum computing and AI ethics must be taken seriously by businesspersons and how these technologies affect strategic decisions; moreover, recommendations and action areas are formulated.

## Introduction

Technology ethics, and moral philosophy in general, include both positive and negative dimensions. That is, ethical approaches aim to bring benefits and avoid causing harm. As a positive example, it may be argued that powerful new technologies which can be used for scientific breakthroughs should be available to many practitioners so that the chances of such breakthroughs are higher and so that people across the world can benefit. As a negative example, it may be argued that biased algorithms which discriminate need to be avoided, for instance to prevent the unjust rejection of a credit application.

While few would disagree with such premises and while there has been analysis of the transformation of business ethics in the information age (De George, 2000), proper awareness of and concrete measures for ethics concerning the emerging fields of artificial intelligence (AI) and quantum technology have not yet been established in most organizations. The fields are naturally considered together as they involve methods to enhance information processing, allowing new insights to be gained more efficiently. There is also a natural symbiosis emerging as quantum computing is being explored to improve machine learning (for instance, increasing its energy efficiency (Cherrat et al., 2024)), and AI approaches, such as artificial neural networks, are being used to enhance quantum technologies (Krenn et al., 2023), for example by reducing the noise plaguing today's quantum computers (Kim et al., 2020).

The discussion of quantum technology here focuses on quantum computing, arguably the one with the greatest transformative potential. Quantum computing and AI will generally be considered together here, often shortened to "quantum AI." It should nevertheless be noted that there are other quantum technologies (in particular, quantum communication and sensing) which are being researched, and there are also "first-generation" quantum technologies that are already widely commercialized in many countries (such as lasers, solid-state electronics and superconductors).

Given the speed at which these technologies have been progressing, it is critical to address ethical questions sooner rather than later. Hence, the key guiding question pursued here is: What are the main arguments for executives, managers and practitioners to take quantum and AI ethics seriously?

AI ethics has already been explored to a degree, including perspectives highlighting the challenges of effectively implementing AI ethics (DG, EPRS 2020; Munn, 2023). Quantum technology ethics, on the other hand, are only starting to come into focus now (Kop, 2021a, 2021b; Quantum Computing Governance Principles 2022; Perrier, 2022; Ménissier, 2022; Possati, 2023; Arrow et al., 2023; Ethics and quantum computing 2024). There is a lack of analysis which combines philosophical arguments and rigor with business considerations. The present work addresses this gap. Six key arguments as to why businesspersons must take quantum AI ethics seriously are shown in Figure 1, approximately ordered by their pertinence and impact.

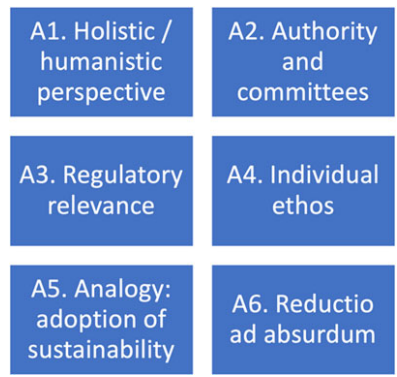

**Figure 1.** Overview of the six main arguments explored in this paper why quantum and AI ethics need to be taken seriously.

The discussion of each argument is structured as follows:

1. Background and relevance to quantum AI
2. Presentation of the argument
3. Example
4. Recommendations

The remainder of the paper considers each argument in turn; following that, a summary is provided, including an overview of overarching action areas.

## Argument from a holistic and humanistic perspective

### Background and relevance to quantum AI

People in business and technology often reduce ethics to what one is allowed or not allowed to do and arguments for and against forbidding or permitting certain actions. Shall it be forbidden that AI systems are used for social scoring, for example? Should the development of more capable and "intelligent" machines be permitted at the risk of potentially spelling an end to the entire human race?

Irrespective of the plausibility of such questions, the exploration of ethics includes many more dimensions. For instance, in the spirit of the ancient Greek philosopher Aristoteles and his ethics of virtue, the field of ethics also deals with questions that concern us personally, such as:

- What is the highest good in life?
- What is it to flourish or live well as a human being?
- What virtues or excellences are needed to flourish and live well?
- How does one develop these virtues or excellences?
- What is friendship and why is friendship a great good?

Another question which is perhaps more in the foreground for the reader is: What is the business value (measured in monetary terms) of pursuing such questions? On the positive side, studies show that meaningful work and identification with work lead to higher productivity (Manyika and Taylor, 2017; Lysova et al., 2019; Nikolova and Cnossen, 2020; Van der Deijl, 2022); this is seen in workplaces where questions about the ethics of virtue are addressed, and meaningful answers are offered. On the negative side, mid-life crises, burnouts, or sick days may follow where those questions are disregarded.

### Presentation of the argument

P(remise) 1: Questions around the ethics of virtue raise awareness about living well and the importance of living well.
P2: If humans are aware of the importance of living well, then it is more likely that those humans will have a good life.
P3: A good life includes meaningful work and identification with work.
P4: If meaningful work and identification with work are absent, then high costs result for companies.

Thus: Engaging in ethics cuts costs for companies.

This is a compelling, yet also very general argument. It applies to a setting where quantum AI is driving transformation but also to many other settings.

The emergence of AI and quantum computing further complicates the already complex meaning–work relationship (Susskind, 2023), including replacing jobs as well as creating new ones (making skill development and transformation topics very important) (Gini 2000; Wulff and Finnestrand, 2023). Even more existentially relevant is the consideration that quantum AI has such transformative potential that "getting it wrong" could literally be humanity's last invention/mistake (Barrat, 2013). Therefore, it is a daunting, or at least very challenging, task for business leaders to spell out the meaning–work relationship in an environment that is increasingly influenced by quantum AI. Attempting to do so is, by its very nature, an ethical endeavor.

### Example

A crucial example in connection with this first argument revolves around responsibility gaps (Kiener, 2022). The general idea behind responsibility gaps is as follows: when technologies take over tasks from human beings, like robots and other AI systems do, there are often worries that there might be cases when the stakes are high and when outcomes might come about for which somebody should be held responsible (both in terms of praise/reward and blame/liability). A case in point is the emergence of self-driving cars (Goodall, 2016). Yet, it might be unclear who, if anybody, could or should be held responsible for a certain positive or negative outcome. Hence, potential responsibility gaps occur.

One domain of work life where there might be reason to be concerned that such responsibility gaps – which can also be called achievement gaps as far as positive outcomes are concerned – might occur is in the domain of hospitals. AI, quantum computing and other advanced technologies are increasingly being introduced into different types of work contexts, where positive responsibilities were previously completely associated with human efforts and ingenuity. Think, for example, of decision-making tools that can help medical doctors diagnose illnesses, for example, by looking at X-ray images of patients and suspected problems they have. AI tools are, at least in certain areas, on the cusp of outperforming human medical doctors both in terms of diagnosing problems and in coming up with treatment regimes. The doctor's role might be reduced to communicating the findings of the AI system to the patient. It might even be reduced further than that because perhaps future AI systems can also come up with highly personalized ways of communicating (medical) information to patients (Flöther et al., 2023).

Such a scenario, which could be a reality very soon, confronts us with a daunting challenge: on the one hand, what is there left to do for highly qualified personnel (knowledge workers) in the era of quantum AI? On the other hand, how will business leaders and managers motivate their colleagues in the future where (large) parts of the colleagues' and specialists' formerly impressive accomplishments (such as analyzing medical images) are attributed to AI systems?

### Recommendations

- Strive to raise awareness among employees about the importance of a balanced sustainable life – the long-term cumulative positive impact of such employees will be greater than those by workaholics that burn themselves out in a short period of time.
- Lay the foundation for a smooth and rewarding human–machine collaboration in the work environment. Articulating an appealing future of work vision is key to winning hearts and minds. In the short term, enhance engagement and identification of employees with whatever task/initiative is at hand. This includes exploring solutions to relieve employees from tedious activities, allowing them to focus on more intriguing work (including applying emerging technology) and enabling them to become life-long learners.
- Develop compelling strategies to close responsibility gaps. Consider fostering teams that are composed of both humans and AI systems. Such teams ought to bear responsibility for the outcomes which are produced by the team. Consider further that responsibility comes with different facets: while the facet of answerability could very well be taken on by artificial systems (e.g., just ask Chat-GPT why it gave you a certain answer), the facet of liability can likely only be meaningfully ascribed to human members of teams composed of both humans and AI systems (imagine suing an AI system or putting it in jail).[1]

## Argument from Authority: Ethics by committee

### Background and relevance to quantum AI

"Ethics by committee/commission" is a popular approach among practitioners who are interested in ethical aspects of their decision-making. This approach consists of assembling a team of usually both inhouse and external experts who are supposed to come up with a set of ethical guidelines that organizations and businesspersons ought to follow (Why You Need an AI Ethics Committee 2023). As it turns out from time to time, however, not every committee's work or not all ethical guidelines are taken seriously. They might have little to no impact on informing actual decision-making of the practitioners that stepped up earlier and made a case for establishing the committee in the first place. In other words, there exists a veritable risk of "ethics washing" if (too) many (industry) people are on the board that have an interest in coming up with vague ethical goals and in avoiding ethics-based regulation, while appearing to care about ethics to strengthen the company's license to operate.[2]

### Presentation of the argument

P1: People A, B, C, … claim that quantum computing and AI ethics are important in business.
P2: People A, B, C, … are experts in quantum AI and business.

Thus: Quantum AI ethics are important to businesspersons.

This is a fallacious argument. If A, B, C claimed that it is important to businesspersons to know the number of certified sommeliers in their organizations, it would also not necessarily follow that this is of actual importance. Yet, arguments from authority and, particularly, ethics by committee are quite common and do have steering power for the practice of quantum AI technology.

AI and particularly quantum technology (Vermaas, 2017) are complex and abstract. Quantum mechanics, with its counterintuitive principles such as superposition and entanglement, is even being used to inspire and further general philosophical discourse in ethics as well as the humanities and social sciences (Voelkner and Zanotti, 2022). Unsurprisingly, therefore, many myths and half-truths exist around quantum computing, propagated by many "authorities". For example:

- Quantum computers will (soon) replace all classical ones;
- Quantum computers will make all calculations faster;
- Quantum computers require a lot of energy;
- Quantum computers are best for (classical) big data problems.

Hence, the constitution, competence and working processes of a quantum AI ethics committee must be even more carefully considered.

### Example

Leveraging internal research as well as published work that illuminates different facets of ethics committees and approaches to the topic (Schrag, 2011), the center of competence for quantum and AI "QuantumBasel" (part of the innovation campus "uptownBasel" in Switzerland) has from the start been exploring the role ethics guidelines and advisors can and should play in an ecosystem which develops cutting-edge technology. In order to foster proactive engagement with ethics and minimize "ethics washing," the approach has encompassed:

- Exploration of guiding principles and their applicability to projects conducted with partner companies;
- Advisors on quantum AI and ethics who have served in other ethics committees and roles and know about the risks around "ethics washing";
- Empowered employees and advisors who proactively raise topics that should be addressed.

### Recommendations

- Articulate clearly how a quantum AI committee and advisors should make decisions – and on what grounds.
- Carry out "postmortem" workshops to proactively address "ethics washing" and related risks.
- Empower the committee and advisors.

## Argument by regulatory relevance

### Background and relevance to quantum AI

Business leaders often care about maximizing profits and not about maximizing the number and impact of ethical actions per se. At least this has been the traditional approach, reflected in the key target of maximizing shareholder value, although recent developments have brought greater attention to aspects such as sustainability and general social responsibility (Denning, 2019).

Still, perhaps there is room for ethics to assist with reaching profit goals. Indeed, ethics are (at least) of instrumental value for companies to develop products which meet regulatory requirements. Engagement with ethics thus turns out to be a competitive advantage, becoming increasingly relevant in a hyperconnected world where brands and reputations are built or impacted within weeks, if not days.

Ethical standards and strong moral beliefs of society members are sooner or later expressed and reflected in regulation. The following risks and insights should be highlighted to clarify the relationship between ethics and regulation further:

- Beware of overregulation.
  ○ Regulation that results in new bureaucracy and potentially unnecessary overhead should not be created hastily. Since such bureaucracy requires financing and triggers follow-up costs in the economy, one needs to carefully consider if sufficient value is created. However, before this question is answered, another one should be addressed. Rationality requires one to first tackle the question of whether to regulate and according to which principles. Both aspects point to the realm of ethics. Only then the "how" becomes relevant, which is of primary concern to legislation.
- Responsibility guidelines are needed first.
  ○ It is not clear which governmental and institutional bodies should be in the lead for which ethical questions. Thus, this must be contemplated before concrete principles and laws can be developed.
- Legislation is restricted to a certain jurisdiction. Ethics, by contrast, can be argued to have universal features.
  ○ As the world is still far away from globally applicable legislation in most fields (if this is desirable at all, given that at least countries with a long tradition of liberalism may be skeptical here), the question arises if ethics-related regulation can be formulated generally enough to be widely adopted without ending up being watered down too much.

Regulation on technology can be severe if society deems it necessary (consider human cloning (Langlois, 2017) or biological weapons (Biological Weapons Convention 2023)). Companies voluntarily pulling out of facial recognition research (Why It Matters That IBM Has Abandoned Its Facial Recognition Technology 2020) and the calls for a global AI moratorium (Miotti and Wasil, 2023) indicate that at least certain aspects of quantum AI are likely to face regulation.

In order to strengthen trust in the responsible use of such emerging technologies, it is very important to sensitize the public to the potential risks of quantum AI applications. Clear principles must be developed to ensure that AI is used responsibly, ethical standards are adhered to and data protection is guaranteed. These include transparency, fairness, security and accountability. Ethical and legal aspects must be taken into account in order to prevent abuse and discrimination. Quantum AI models reflect the data with which they are trained and the people who program them. It is crucial that existing biases and discrimination in society are not reproduced.

Furthermore, if quantum AI really does live up to its lofty expectations, this will bring the question into the foreground concerning who has access to such powerful technology. AI, including large language models and foundation models, already cost millions, and soon perhaps billions, of US dollars to train (The cost of training AI could soon become too much to bear 2024). Access to quantum computers is (still) relatively limited, and there are significant disparities between different regions (State of Compute Access: How to Bridge the New Digital Divide 2023). Organizations or countries who are not leaders in this space may argue that they should not be excluded from the positive benefits of breakthroughs, for instance longer life expectancies due to the design of novel drugs.

*Presentation of the argument*

P1: Moral beliefs in society shape tomorrow's regulation.
P2: Ethics is the branch of philosophy which deals with questions of human morality.
P3: Items of regulatory importance are of practical importance to businesspersons.

Thus: Ethics is important to businesspersons.

Premises 1 and 2 are straightforward. Premise 3 can be rationalized once one recalls that regulation decides on what defines a market-compliant product which businesses aim to sell. In this way, engaging with ethics increases revenue or at least prevents revenue loss (The Business Risks of Poor AI Ethics and Governance 2022).

*Example*

A variety of guidelines has been issued across the world around ethical AI (Jobin et al., 2019). While the precise relationship between ethics and AI legislation continues to stir debate (Anderson, 2022; Catanzariti, 2023), governments are moving from AI guidelines toward binding regulation now. For example, the EU AI Act represents one of the world's first efforts to regulate AI (EU agrees "historic" deal with world's first laws to regulate AI 2023). It is designed to help AI uptake while banning risky AI technology and applications (Schuett, 2023). Such legislation will significantly influence how businesses engineer and apply AI models.

One contentious point in the act concerns the development and application of foundation models; (ethical) arguments around regulating foundation models include their use in the development of general-purpose AI. However, restriction of foundation models would arguably limit the competitiveness of EU companies compared with the rest of the world (Will Disagreement Over Foundation Models Put the EU AI Act at Risk? 2023). Hence, the ethics surrounding (quantum) AI regulation have a direct impact on businesspersons.

*Recommendations*

- Be clear about which primary targets your business has – shareholder value? Ethics? Social goals? Long-term survival?
- Be flexible and adapt as new laws loom based on shifts in technological progress and societal values.
- Be mindful of international differences in regulation.

**Argument for acknowledging complexity: the case for individual ethos, regulation is not enough**

*Background and relevance to quantum AI*

The picture that quantum and AI industries and practices present us with is not black and white, but rather complex, both from a descriptive and from a normative perspective. The focus of reform efforts toward responsible and trustworthy AI has been on legal regulation: changes in the rules and regulations that govern AI systems. While doubtlessly important, however, one wonders whether such changes in legal regulation are sufficient or whether they need to be accompanied by changes in the ethos and the culture of AI and quantum-related markets.

In systems as complex as today's economic systems, governance by rules requires getting the incentives right, which is no easy task. Rules are, by definition, rather rigid and may not be ideal as the only tool for regulating a wide variety of cases. If circumstances change quickly, for instance due to the rapid and often not-easily-understood advances and developments of quantum AI, rules may become quickly outdated and those setting them may have trouble catching up. Rules need to be applied to concrete cases, which can require some degree of judgment and a joint understanding of the practices they are supposed to regulate.

Furthermore, the control of whether rules are obeyed is time-consuming and costly. These various factors imply that it is highly desirable that complex systems be regulated not exclusively by rules but also by a joint ethos of those who participate in the systems – a professional ethos that embodies an orientation toward the standards implicit in the practices they are active in.[3]

Quantum AI exacerbates these issues and can easily result in black box models and approaches. For example, models developed with deep neural networks are typically difficult for humans to comprehend, and their outputs are not easily explainable. However, it has been argued that a "right to explanation" is present when algorithmic decision-making is involved (Kim and Routledge, 2022). Quantum computing, by its very nature, is probabilistic; this may create replicability challenges, further complicated by the noise affecting current generations of quantum systems. Thus, the complexity and pace associated with the technology mean that individual ethos must always accompany general regulation.

### *Presentation of the argument*

P1: Quantum AI technology is complex and embedded in complex social systems.
P2: Complex systems cannot be steered and governed by regulation and rigid rules only.
P3: If regulation and rigid rules are insufficient for steering and governing complex systems, then there is a strong need for individual ethos and ethics.

Thus: Quantum AI systems require ethos and ethics from developers and users so that they are employed in a way which is desired by society.

### *Example*

It has been estimated that insiders cause 20% of security breaches (DBIR: Data Breach Investigations Report 2008-2022). However, security budgets are primarily allocated to external threats (Cybercrime Is An Inside Job 2020); individual care and ethos tends to be neglected. This imbalance is reflective of the challenge of containing a budding technology such as quantum AI. Even the best (external) regulation will not be able to prevent individuals from crossing red lines if they lack the ethos that prevents them from doing so.

### *Recommendations*

- Inculcate employees, be they managers, scientists, or developers, with awareness around the potential impact of quantum AI technology and their individual responsibilities.
- Anticipate developments and plan for appropriate responses, for example through scenario analysis and postmortem exercises.
- Achieve clarity around individual incentives and motivation – and address these proactively, particularly should they disincentivize individual ethos.

## Argument by analogy: The case of sustainability

### *Background and relevance to quantum AI*

Sustainability, as a normative concept and social goal for people (including future generations (MacAskill, 2022)) to co-exist on Earth over a long time along typically the three dimensions of the environment, economy and society, has been a megatrend for some years if not decades.[4] Prominently, the United Nations agreed the Sustainable Development Goals (SDGs) in 2015, which set a global agenda for sustainable development, with a deadline of 2030.

For our purposes, the attention toward sustainability from the business practice is remarkable in this context. To give just three examples, this large and growing attention is documented by 1) a variety of clean-tech, social impact and food-tech startups nowadays that respond to the SDGs, 2) novel business and policy concepts such as carbon credits and 3) the sheer number of sustainability reports that have been voluntarily published by many companies throughout different sectors. This, in turn, has spun the wheel of a flourishing ecosystem through developments that include, for instance, metrics such as Bloomberg's environmental and social governance (ESG) ratings and the Dow Jones Sustainability Indices (DSJI), which use private and public information from companies, and serve as proxies for actual sustainability performance. Another example is the formation and growth of organizations such as the Ellen MacArthur Foundation and the World Business Council for Sustainable Development.

To make a case for ethics in quantum AI markets, one can take inspiration from how sustainability has transitioned from an academic topic to a policymaker issue to a theme of interest for most companies, large or small. The idea is that one compares where quantum AI stands today with where sustainability was years/decades ago. One can then use this comparison for predictions of quantum AI development and adoption.

### *Presentation of the argument*

P1: Sustainability has transitioned from a policymaker topic to a theme of interest for most companies.
P2: Sustainability and quantum AI ethics are sufficiently similar, for instance in respect to their cross-societal impact and close connections to the future of human wellbeing.

Thus: Quantum AI ethics will transition from a policymaker topic to a theme of interest for most companies.

### *Example*

Sustainability has been argued to be indispensable to corporate strategy (Why sustainability is crucial for corporate strategy 2022)

and to have a clear business case (Whelan and Fink, 2016). Moreover, early movers have already reaped rewards, an example being organic pioneers who enjoyed greater price premiums before the organic produce market significantly increased (Early movers in sustainability reap rewards 2020). Companies are increasingly considering sustainability and AI together now in order to achieve competitive advantages and further social objectives, cases in point being Microsoft and GE Research (The Opportunities at the Intersection of AI, Sustainability, and Project Management 2023). Quantum computing, in addition, is also being explored to help achieve climate goals (Cooper et al., 2022).

Moreover, sustainability and digital technologies are becoming more and more closely intertwined; key challenges no longer just relate to digital innovation but increasingly also to governance, bringing together "the green and the blue" (Floridi 2020). This intimate connection between the green and the blue is intricate and multilayered. For instance, while AI systems can make processes more efficient (which saves resources and cuts emissions), their development and training also increase carbon emissions; thus, a system of systems view is required (Gaur et al., 2023). This includes quantum computing, which is increasingly being explored to improve the energy efficiency of computing, including AI applications (Chen, 2023).

### Recommendations

Learn from the emergence of sustainability and other transitions and avoid:

- Ignoring the topic until it is "too late";
- Not proactively developing a perspective on how quantum AI can be used for good;
- Not reaping early-mover rewards, such as positive branding and novel business models.

## Reductio ad absurdum: Argument by assuming the opposite scenario leading to unacceptable consequences

### Background and relevance to quantum AI

Good predictions, such as on the future of quantum AI, crop up in tandem with good explanations (Toulmin, 1961). Often something more than mere prediction is desired. One needs to have information about the underlying mechanisms in order to make accurate and robust predictions about what will happen when the system such as a certain market breaks down or modifies itself in various ways in response to disrupting new technologies that are penetrating it. To embark on this arduous journey, humans in the business practice and elsewhere are often counterfactual learners, that is, they imagine worlds that do not exist and infer reasons for observed phenomena. "As-if" exploratory modes and imagination should not be frowned upon but rather treated as an integral substrate of progress (Hoffmann, 2022).

The sixth argument pays tribute to this insight by imagining a possible world where one, as a businessperson or company, decides not to consider ethical aspects of quantum AI. What would be the consequences of this decision? It is argued here that the consequences could easily be severe and unacceptable to good and reflected business leaders since a number of disastrous risks could materialize, such as:

- Losing reputation and damaging the brand;
- Steering into regulatory difficulties in the near future;
- Failing to attract young talents who wish to improve the world;
- Having one's license to operate withdrawn by society.

Consider the fourth point: younger generations no longer choose employment based on a small set of factors such as salary and manager. They expect to work on cutting-edge problems with a strong positive impact. Quantum AI presents such opportunities par excellence because it is in the transition from lab to application (meaning many interesting research questions must still be addressed, and there are many white spaces to explore) and because it has tremendous transformative potential; quantum AI even raises fundamental questions about the nature of reality – for example, it has been argued that the mere existence of quantum computers provides evidence for a multiverse (Deutsch, 1997; Tegmark, 2003).

Given quantum AI's complexity coupled with its revolutionary potential, "getting it wrong" could easily lead to operating a business that is not in line with society's zeitgeist and thus considerable reputation and regulation problems. For instance, quantum computing has significant potential in military applications (Krelina, 2021) and, as with any technology, could also be used for nefarious goals. A prominent example here is in cybersecurity and data privacy where many current cryptographic standards are being threatened by future generations of quantum computers ("harvest now, decrypt later" attacks further highlight the importance of considering such issues already today).

### Presentation of the argument

P1: Ethical aspects of quantum AI are irrelevant to businesspersons.
P2: If ethical aspects of quantum AI are irrelevant to businesspersons, then costs follow for businesspersons.
P3: Businesspersons aim to avoid costs.

Thus: Ethical aspects of quantum AI are not irrelevant to businesspersons. In other words, the risk is *not* to be in quantum; the risk is *not to be* in quantum.

### Example

The Facebook-Cambridge Analytica data scandal, where personal data from millions of Facebook users was used via machine learning for political advertising that arguably had a significant influence on political outcomes, led to the greatest crisis for Facebook in its history and to Cambridge Analytica ceasing its operations (The Cambridge Analytica scandal changed the world – but it didn't change Facebook 2019). It highlights the above risks and potential consequences for businesses who do not explore quantum AI ethics with sufficient care.

### Recommendations

- Consider if the risks to your business of not exploring quantum AI compared with the risks of exploring quantum AI.
- Estimate the impact of certain scenarios (for instance, failure to attract quantum talent or failure to own intellectual property rights for a key quantum AI application) and use these to guide prioritization of quantum AI initiatives.
- Continually reassess – quantum AI technology is developing so quickly that plausible future scenarios today may already be replaced by new ones tomorrow.

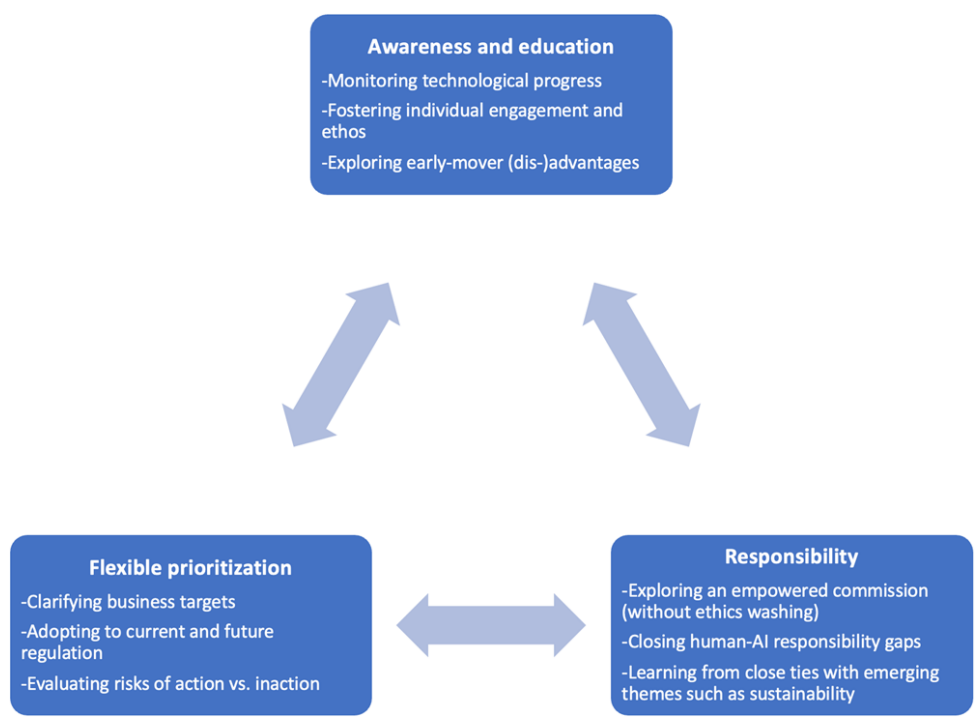

Figure 2. Overview of action areas for the exploration and adoption of quantum AI ethics.

## Summary and action areas

The last years have seen drastic progress in the development and adoption of AI and quantum computing. While the "holy grail" for both technologies is likely still years or decades away, namely artificial general intelligence and large-scale fault-tolerant quantum computation, more and more hurdles are cleared en route to those targets (Bubeck et al., 2023; Bluvstein et al., 2023). Increasingly, the technologies are intersecting, and a natural symbiosis is developing where progress in one accelerates progress in the other. This rapid change by and through quantum AI, not even accounting for the emergence of other technologies that may drive further synergies such as neuromorphic computing (Marković et al., 2020), increases the urgency with which ethical considerations surrounding both technologies must be addressed. Far from being solely an esoteric academic question, quantum AI ethics are of great importance to many. The present paper contains six key arguments why businesspersons must take quantum AI ethics seriously today as well as recommendations and best practices for each argument area.

As is (almost) always the case, "little and often" tends to beat "rarely but strongly." Therefore, it is likely neither advisable to create a large temporary team that conducts a quantum AI (ethics) deep dive and then stops the work after submitting a lengthy report nor, of course, to remain idle and do nothing. For the latter option, while action always carries opportunity costs, the risks of inaction are great. The speed of the technological development continually creates new intricacies and raises fresh questions that must be monitored and considered in strategic choices.

The wide range of ethical issues associated with quantum AI demands also varied measures and actions. The recommendations from this paper are abstracted into overarching action areas in Figure 2.

Ultimately, quantum AI ethics requires considerably more research and social discourse. Businesspersons, together with researchers and the rest of society, should strive to set a positive example by being proactive – and thus helping guide industries, economies and the entire world toward prosperity in the age of AI and quantum technology.

## Notes

1 Ascribing responsibility (in a certain sense: answerability vs. liability) to non-biological agents is less radical and novel than it looks like at first glance. For example, corporations are non-human entities, just like AI systems, and they bear responsibility, which is the well-established subject matter of Corporate Responsibility (CR)/Corporate Social Responsibility (CSR).
2 The license to operate is an established term from business ethics and refers to the social acceptance of companies. This is based on the intersubjective perception of members of society and, therefore, cannot be formally acquired. In the wake of growing criticism of companies and their value creation, the question of how to safeguard the license to operate is becoming increasingly relevant. Lack of a license to operate manifests itself in a gradual loss of the entrepreneurial ability to cooperate, as a result of which entrepreneurial value creation becomes more difficult; it can therefore also be described as the basis of entrepreneurial value creation.
3 One would want AI and quantum professionals to organize themselves in professional bodies in which they can discuss, and reinforce, ethical norms; such professional bodies might also take on legal liability for certain kinds of risks or harms. One would want them to build and maintain an ethos that is appropriate to the responsibility of an industry which plays a crucial role in modern societies. But since, admittedly, such an ethos may be elusive for now, more thoughts and guidance are required at this stage.
4 The 1983 UN Commission on Environment and Development (Brundtland Commission), defining "sustainability" in their report as a development that "meets the needs of the present without compromising the ability of future generations to meet their own needs," has had a major influence on sustainability discourses.

**Data availability statement.** Data availability is not applicable to this article as no new data were created or analyzed in this study.

**Acknowledgments.** The authors would like to thank Yuval Boger, Stephen Cave, Maria Fay, Martina Gschwendtner, Matthias Hölling, Hans Noser, Henning Soller, Lothar Thiele and Michael Tschudin for helpful discussions.

**Financial support.** This research received no specific grant from any funding agency or commercial/not-for-profit sectors.

**Competing interests.** The authors declare no conflicts of interest.

**Ethics statement.** Ethical approval and consent are not relevant to this article.

## Connections references

**D'Auria V and Teller M** (2023) What are the priorities and the points to be addressed by a legal framework for quantum technologies? *Research Directions: Quantum Technologies* **1**, e9. https://doi.org/10.1017/qut.2023.3.

## References

**Anderson MM** (2022) Some ethical reflections on the EU AI Act. In *IAIL 2022: 1st International Workshop on Imagining the AI Landscape After the AI Act, vol. 3221.*

**Arrow JÉ., Marsh SE and Meyer JC** (2023) A holistic approach to quantum ethics education: The quantum ethics project quantumethicsproject.org. (*2023 IEEE International Conference on Quantum Computing and Engineering (QCE), vol. 3*, IEEE, pp. 119–128,

**Barrat J** (2013) *Our Final Invention: Artificial Intelligence and the End of the Human Era.* United States: St. Martin's Press.

**Biological Weapons Convention.** United Nations Office for Disarmament Affairs. Available at https://disarmament.unoda.org/biological-weapons, (accessed November 29, 2023).

**Bluvstein D, Evered SJ, Geim AA, Li Sophie H, Zhou H, Manovitz T, Ebadi S, Cain M, Kalinowski M, Hangleiter D, Bonilla Ataides JP, Maskara N, Cong I, Gao X, Sales Rodriguez P, Karolyshyn T, Semeghini G, Gullans MJ, Greiner M, Vuletić V and Lukin MD** (2023) Logical quantum processor based on reconfigurable atom arrays. *Nature* **626**(7997), 1–3.

**Bubeck S, Chandrasekaran V, Eldan R, Gehrke J, Horvitz E, Kamar E, Lee P, et al.** (2023) Sparks of artificial general intelligence: early experiments with gpt-4, arXiv preprint arXiv:2303.12712.

**Catanzariti M** (2023) What role for ethics in the law of AI?. In *Artificial Intelligence, Social Harms and Human Rights*. Cham: Springer International Publishing, pp. 141–159.

**Chen S** (2023) Are quantum computers really energy efficient? *Nature Computational Science* **3**(6), 457–460. https://doi.org/10.1038/s43588-023-00459-6.

**Cherrat EA, Kerenidis I, Mathur N, Landman J, Strahm M and Yvonna Li Y** (2024) Quantum vision transformers. *Quantum* **8**, 1265. https://doi.org/10.22331/q.

**Cooper P, Ernst P, Kiewell D and Pinner D** (2022) *Quantum Computing Just Might Save the Planet*. McKinsey Digital. https://www.mckinsey.com/capabilities/mckinsey-digital/our-insights/quantum-computing-just-might-save-the-planet.

Cybercrime Is an inside job. *Cybercrime Magazine.* https://cybersecurityventures.com/cybercrime-is-an-inside-job (last modified November 16, 2020).

**DBIR: Data Breach Investigations Report** (2008-2022), https://www.verizon.com/business/resources/T202/reports/dbir/2022-data-breach-investigations-report-dbir.pdf (last modified September 23, 2022).

**De George RT** (2000) Business ethics and the challenge of the information age. *Business Ethics Quarterly* **10**(1), 63–72. https://doi.org/10.2307/3857695.

**Denning S** (2019) *Why Maximizing Shareholder Value Is Finally Dying*, Forbes.

**Deutsch D** (1997) *the Fabric of Reality*. London, UK: Allen Lane Penguin Press N. https://www.penguin.co.uk/.

**DG, EPRS** (2020) The ethics of artificial intelligence: issues and initiatives.

Early movers in sustainability reap rewards, *The Packer*. https://www.thepacker.com/news/sustainability/early-movers-sustainability-reap-rewards, (last modified September 24, 2020).

Ethics and quantum computing. *Scientific Computing World*. https://www.scientific-computing.com/article/ethics-quantum-computing (accessed February 1, 2024).

EU agrees 'historic' deal with world's first laws to regulate AI, *The Guardian*. https://www.theguardian.com/world/2023/dec/08/eu-agrees-historic-deal-with-worlds-first-laws-to-regulate-ai (last modified December 9, 2023).

**Floridi L** (2020) The green and the blue: a new political ontology for a mature information society, Available at SSRN 3831094.

**Flöther FF, Kwatra S, Lustenberger P and Ravizza S.** Communication content tailoring, U.S. Patent US11641330B2, filed 6 August 2020, and issued 2 May 2023.

**Gaur L, Afaq A, Kaur Arora G and Khan N** (2023) Artificial intelligence for carbon emissions using system of systems theory. *Ecological Informatics* **76**, 102165. https://doi.org/10.1016/j.ecoinf.2023.102165.

**Gini A** (2000) What happens if work goes away? *Business Ethics Quarterly* **10**(1), 181–188. https://doi.org/10.2307/3857704.

**Goodall NJ** (2016) Can you program ethics into a self-driving car? *IEEE Spectrum* **53**(6), 28–58. https://doi.org/10.1109/MSPEC.2016.7473149.

**Hoffmann CH** (2022) *The Quest for a Universal Theory of Intelligence: The Mind, the Machine, and Singularity Hypotheses*, Walter de Gruyter GmbH & Co KG.

**Jobin A, Ienca M and Vayena E** (2019) The global landscape of AI ethics guidelines. *Nature Machine Intelligence* **1**(9), 389–399. https://doi.org/10.1038/s42256-019-0088-2.

**Kiener M** (2022) Can we bridge AI's responsibility gap at will? *Ethical Theory and Moral Practice* **25**(4), 575–593. https://doi.org/10.1007/s10677-022-10313-9.

**Kim C, Daniel Park K and Rhee J-K** (2020) Quantum error mitigation with artificial neural network. *IEEE Access* **8**, 188853–188860. https://doi.org/10.1109/Access.6287639.

**Kim TW and Routledge BR** (2022) Why a right to an explanation of algorithmic decision-making should exist: a trust-based approach. *Business Ethics Quarterly* **32**(1), 75–102. https://doi.org/10.1017/beq.2021.3.

**Kop M** (2021a) *Why We Need to Consider the Ethical Implications of Quantum Technologies*. Physics World. https://physicsworld.com/a/why-we-need-to-consider-the-ethical-implications-of-quantum-technologies/.

**Kop M** (2021b) Establishing a legal-ethical framework for quantum technology. *Yale Law School, Yale Journal of Law & Technology (YJoLT), The Record*

**Krelina M** (2021) Quantum technology for military applications. *EPJ Quantum Technology* **8**(1), 24. https://doi.org/10.1140/epjqt/s40507-021-00113-y.

**Krenn M, Landgraf J, Foesel T and Marquardt F** (2023) Artificial intelligence and machine learning for quantum technologies. *Physical Review A* **107**(1), 010101. https://doi.org/10.1103/PhysRevA.107.010101.

**Langlois A** (2017) The global governance of human cloning: the case of UNESCO. *Palgrave Communications* **3**(1), 1–8. https://doi.org/10.1057/palcomms.2017.19.

**Lysova EI, Allan BA, Dik BJ, Duffy RD and Steger MF** (2019) Fostering meaningful work in organizations: a multi-level review and integration. *Journal of Vocational Behavior* **110**, 374–389. https://doi.org/10.1016/j.jvb.2018.07.004.

**MacAskill W** (2022) *What We Owe the Future*. New York: Basic Books. https://www.hachettebookgroup.com/imprint/basic-books/.

**Manyika J and Taylor M** (2017) *Creating Meaningful Work and Driving Business Success Amid Technological Disruption*. McKinsey Global Institute. https://www.mckinsey.com/featured-insights/future-of-work/creating-meaningful-work-and-driving-business-success-amid-technological-disruption.

**Marković D, Mizrahi A, Querlioz D and Grollier J** (2020) Physics for neuromorphic computing. *Nature Reviews Physics* **2**(9), 499–510. https://doi.org/10.1038/s42254-020-0208-2.

**Ménissier T** (2022) The ethics of quantum technologies: a problematic formulation, In *QuantAlps Days*.

**Miotti A and Wasil A** (2023) An international treaty to implement a global compute cap for advanced artificial intelligence. arXiv preprint arXiv:2311.10748.

**Munn L** (2023) The uselessness of AI ethics. *AI and Ethics* **3**(3), 869–877. https://doi.org/10.1007/s43681-022-00209-w.

**Nikolova M and Cnossen F** (2020) What makes work meaningful and why economists should care about it. *Labour Economics* **65**, 101847. https://doi.org/10.1016/j.labeco.2020.101847.

**Perrier E** (2022) Ethical quantum computing: a roadmap, arXiv preprint arXiv: 2102.00759.

**Possati LM** (2023) Ethics of quantum computing: an outline. *Philosophy & Technology* **36**, 48. https://doi.org/10.1007/s13347-023-00651-6.

**Schrag ZM** (2011) The case against ethics review in the social sciences. *Research Ethics* **7**(4), 120–131. https://doi.org/10.1177/174701611100700402.

**Schuett J** (2023) Risk management in the artificial intelligence act. *European Journal of Risk Regulation* 1–19. https://www.cambridge.org/core/services/aop-cambridge-core/content/view/2E4D5707E65EFB3251A76E288BA74068/S1867299X23000016a.pdf/risk-management-in-the-artificial-intelligence-act.pdf.

State of compute access: how to bridge the new digital divide, *Tony Blair Institute for Global Change*. https://www.institute.global/insights/tech-and-digitalisation/state-of-compute-access-how-to-bridge-the-new-digital-divide (last modified December 7, 2023).

**Susskind D** (2023) Work and meaning in the age of AI.

**Tegmark M** (2003) Parallel universes. *Scientific American* **288**(5), 40–51. https://doi.org/10.1038/scientificamerican0503-40.

The Business Risks of Poor AI Ethics and Governance, Arek Skuza Innovation and Growth. https://arekskuza.com/the-innovation-blog/ai-ethics (last modified June 13, 2022).

The Cambridge Analytica scandal changed the world – but it didn't change Facebook, *The Guardian*. https://www.thepacker.com/news/sustainability/early-movers-sustainability-reap-rewards (last modified March 18, 2019).

The cost of training AI could soon become too much to bear, *Fortune*. https://fortune.com/2024/04/04/ai-training-costs-how-much-is-too-much-openai-gpt-anthropic-microsoft/ (last modified April 4, 2024).

The opportunities at the intersection of AI, sustainability, and project management, *Harvard Business Review*. https://hbr.org/2023/10/the-opportunities-at-the-intersection-of-ai-sustainability-and-project-management, (last modified October 27, 2023).

**Toulmin SE** (1961) *Foresight and Understanding: An Enquiry into the Aims of Science*. New York: Harper & Row.

**Van der Deijl W** (2022) Two concepts of meaningful work. *Journal of Applied Philosophy* **41**. https://onlinelibrary.wiley.com/doi/epdf/10.1111/japp.12614.

**Vermaas PE** (2017) The societal impact of the emerging quantum technologies: a renewed urgency to make quantum theory understandable. *Ethics and Information Technology* **19**(4), 241–246. https://doi.org/10.1007/s10676-017-9429-1.

**Voelkner N and Zanotti L** (2022) Ethics in a quantum world. *Global Studies Quarterly* **2**(3), 1–5. https://academic.oup.com/isagsq/article/2/3/ksac044/6717714.

**Whelan T and Fink C** (2016) The comprehensive business case for sustainability. *Harvard Business Review* **21**. https://sandiego.ascm.org/images/downloads/Eco_Sustainability/the_comprehensive_business_case_for_sustainability.pdf.

Why it matters that IBM has abandoned its facial recognition technology, *Forbes*. https://www.forbes.com/sites/timbajarin/2020/06/18/why-it-matters-that-ibm-has-abandoned-its-facial-recognition-technology (last modified June 18, 2020).

Why you need an AI ethics committee, *Harvard Business Review*. https://hbr.org/2022/07/why-you-need-an-ai-ethics-committee (accessed November 29, 2023).

Will Disagreement Over Foundation Models Put the EU AI Act at Risk?, Tech Policy Press. https://techpolicy.press/will-disagreement-over-foundation-models-put-the-eu-ai-act-at-risk (last modified November).

**World Economic Forum**. Quantum Computing Governance Principles. https://www3.weforum.org/docs/WEF_Quantum_Computing (last modified January, 2022).

**World Economic Forum**. Why Sustainability is Crucial for Corporate Strategy. https://www.weforum.org/agenda/2022/06/why-sustainability-is-crucial-for-corporate-strategy (last modified June 9, 2022).

**Wulff K and Finnestrand H** (2023) Creating meaningful work in the age of AI: explainable AI, explainability, and why it matters to organizational designers. *AI & Society* **39**, 1–14. https://link.springer.com/article/10.1007/s00146-023-01633-0.